\documentclass[]{raa}

\usepackage{color}
\usepackage{graphicx}
\usepackage{times}
\usepackage{natbib}
\usepackage{subcaption}
\usepackage{amssymb,amsmath}
\usepackage[pagebackref=true]{hyperref}

\bibpunct{(}{)}{;}{a}{}{,}

\begin{document}

  \title{Cross-correlation Forecast of CSST Spectroscopic Galaxy and MeerKAT Neutral Hydrogen Intensity Mapping Surveys}
  \volnopage{Vol.0 (20xx) No.0, 000--000}      
   \setcounter{page}{1}          

   \author{Yu-Er Jiang \inst{1,2}
   \and Yan Gong \inst{1,3,4*}
   \and Meng Zhang \inst{1,2}
   \and Qi Xiong \inst{1,2}
   \and Xingchen Zhou \inst{1,2}
   \and Furen Deng \inst{1,2}
   \and Xuelei Chen \inst{1,2,5,6}
   \and Yin-Zhe Ma \inst{4,7,8}
   \and Bin Yue \inst{1}
   }

   \institute{National Astronomical Observatories, Chinese Academy of Sciences, Beijing 100012, China; {\it gongyan@bao.ac.cn}\\
        \and
            School of Astronomy and Space Sciences, University of Chinese Academy of Science (UCAS), Beijing 100049, China \\
        \and
            Science Center for China Space Station Telescope, National Astronomical Observatories, Chinese Academy of Sciences, Beijing 100101, China\\
        \and
            NAOC-UKZN Computational Astrophysics Centre (NUCAC), University of KwaZulu-Natal, Durban, 4000, South Africa \\
        \and
            Department of Physics, College of Sciences, Northeastern University, Shenyang 110819, China \\
        \and
            Center for High Energy Physics, Peking University, Beijing 100871, China \\
        \and
            School of Chemistry and Physics, University of KwaZulu-Natal, Westville Campus, Private Bag X54001, Durban 4000, South Africa \\
        \and
            Department of Physics, Stellenbosch University, Matieland 7602, South Africa \\
    \vs\no
   {\small Received 20xx month day; accepted 20xx month day}}

\abstract{Cross-correlating the data of neutral hydrogen (H\textsc{i}) 21cm intensity mapping with galaxy surveys is an effective method to extract astrophysical and cosmological information. In this work, we investigate the cross-correlation of MeerKAT single-dish mode H\textsc{i} intensity mapping and China Space Station Telescope (CSST) spectroscopic galaxy surveys. We simulate a survey area of $\sim 300$ $\mathrm{deg}^2$ of MeerKAT and CSST surveys at $z=0.5$ using Multi-Dark $N$-body simulation. The PCA algorithm is applied to remove the foregrounds of H\textsc{i} intensity mapping, and signal compensation is considered to solve the signal loss problem in H\textsc{i}-galaxy cross power spectrum caused by the foreground removal process. We find that from CSST galaxy auto and MeerKAT-CSST cross power spectra, the constraint accuracy of the parameter product $\Omega_{\rm H\textsc{i}}b_{\rm H\textsc{i}}r_{{\rm H\textsc{i}},g}$ can reach $\sim1\%$, which is about one order of magnitude higher than the current results. After performing the full MeerKAT H\textsc{i} intensity mapping survey with 5000 deg$^2$ survey area, the accuracy can be enhanced to $<0.3\%$. This implies that the MeerKAT-CSST cross-correlation can be a powerful tool to probe the cosmic H\textsc{i} property and the evolution of galaxies and the Universe.
\keywords{intensity mapping, large-scale structure, cosmological constraint}
}

   \authorrunning{Y.-E. Jiang, et al. }            
   \titlerunning{Cross-correlation forecast of CSS-OS and MeerKAT IM}  

   \maketitle

\section{Introduction}\label{sec:intro}

Probing the large-scale structure (LSS) of the Universe has always been one of the main missions of cosmological observations. Constraining the property of dark matter and dark energy, recovering the primordial fluctuations and testing gravity theories are all in need of cosmological  surveys with large survey area and wide redshift coverage. To achieve this target, line intensity mapping (LIM) has been proposed and proven to be an efficient technique. LIM makes use of the emission lines from the energy level transition of atoms or molecules, such as H\textsc{i} 21cm, C\textsc{ii}, CO, Ly$\alpha$, H$\alpha$, [O\textsc{iii}], etc.~\citep[see e.g.][]{intro_2010_Visbal,intro_2011_Carilli,intro_2011_Lidz,intro_2011_Gong,intro_2012_Gong,intro_2013_Gong,intro_2014_Gong,intro_2013_Silva,intro_2015_Silva,intro_2014_Pullen,intro_2014_Uzgil,Gong2017,intro_2017_Fonseca,intro_2020_Gong}.
These lines can reflect different properties and progresses of galaxy evolution, and can be good tracers of the LSS. 

Instead of the traditional observations targeting the resolvable sources, intensity mapping probes accumulative intensity of all sources in a spatial volume (voxel) defined by survey spatial and frequency resolutions. So even though some sources are too faint to be detected in traditional sky surveys, in principle, their signals can be probed in intensity mapping. In addition, the frequency shifts of the emission lines are the natural probe of redshift, so intensity mapping is expected to be a powerful tool to obtain cosmic 3D matter structure information traced by emission lines from galaxies with high efficiency and relatively low cost. Among various emission lines, H\textsc{i} 21cm line from atomic hydrogen is the most widely studied in intensity mapping research~\citep[see e.g.][]{HI_Tianlai1, HI_Tianlai2, HI_Bingo1, HI_Bingo2,HI_CHIME1, HI_CHIME2, HI_SKA1, HI_FAST,Meerkat_temperature, MeerKAT&WiggleZ, DFR,
   HI_Bingo3, HI_SKA2,HI_Tianlai3}. Besides being a main probe of epoch of reionization, the neutral hydrogen 21cm line has a tight connection with star formation and galaxy evolution, and it can trace the galaxy and hence dark matter distribution at low and high redshifts.

While many experiments about H\textsc{i} intensity mapping have been proposed or are already running, the foreground contamination problem is still one of the biggest challenges, as the foregrounds can be as large as five orders of magnitude higher than the signal. The high brightness temperature of the Galactic emission and other sources makes H\textsc{i} signal hard to be detected from auto-correlations. In principle, cross-correlating the 21cm observation with an optical galaxy survey in the same survey area is a good method to reduce the foreground contamination and instrumental noise, and extract the signal \citep[e.g.][]{GBT&Deep2}. The signal-to-noise ratio (SNR) can be significantly improved since the foregrounds and instrumental noise of different wave bands in different surveys are barely correlated. 

However, in practice, the cross-correlation result is not fully satisfied due to the complex components of the foreground. So the foreground removal algorithms are still needed in cross-correlations. Various algorithms have been applied, including the blind foreground removal techniques like principal component analysis (PCA) \citep{PCA1} and independent component analysis (ICA) \citep{ICA1} which make use of different frequency smoothness of foreground and signal, the polynomial/parametric-fitting method which fits the physical properties of the foreground \citep{fitting1}, and machine learning (ML) methods \citep{ML}, etc. Although signal loss and foreground residual are usually inevitable, foreground removal techniques do make progress and are necessary in cross-correlation detection.


Currently, positive results on H\textsc{i} abundance and H\textsc{i}-galaxy correlation have been obtained by several experiments. The Green Bank Telescope (GBT) has implemented their H\textsc{i} intensity mapping correlation detection with the Deep2 optical redshift survey \citep{GBT&Deep2}, WiggleZ Dark Energy Survey \citep{GBT&WiggleZ} and eBOSS survey \citep{GBT&eBOSS}. 
In addition, the Parkes radio telescope also presented their work on correlating H\textsc{i} intensity mapping with the 2dF galaxy survey \citep{Parkes&2dF}. 
Recently, MeerKAT accomplished H\textsc{i} intensity mapping correlation detection with the WiggleZ survey \citep{MeerKAT&WiggleZ}. 
They all constrain the H\textsc{i}-galaxy correlation parameter product $\Omega_{\rm H\textsc{i}}b_{\rm H\textsc{i}}r_{{\rm H\textsc{i},}g}$ at different redshifts, where $\Omega_{\rm H\textsc{i}}$, $b_{\rm H\textsc{i}}$, and $r_{{\rm H\textsc{i},}g}$ are the H\textsc{i} energy density parameter, H\textsc{i} bias, and correlation coefficient of H\textsc{i} and galaxy, respectively. 
In this work, we will determine the constraint power on neutral hydrogen parameters by the observations of MeerKAT and the next-generation galaxy survey of China Space Station Telescope (CSST).

MeerKAT is a pathfinder project of the Square Kilometre Array (SKA) and in the future will become a part of SKA-mid \citep{MeerKAT,SKA}. 
It is a state-of-the-art intensity mapping instrument which is capable of complementing and extending cosmological measurements at a wide range of wavelengths. While MeerKAT is a large interferometric array which can access small scales of cosmic structure, single-dish mode is preferred in intensity mapping experiments. We plan to perform MeerKAT H\textsc{i} intensity mapping cross-correlation with the China Space Station Optical Survey (CSS-OS) \citep{Zhan2011,Zhan2021,CSS-OS2,CSS_OS}.
CSS-OS is the major observation project of CSST, and it will cover 17,500 $\mathrm{deg}^2$ of sky area in a 10 yr working time. In addition, the spectroscopic survey of CSS-OS will provide a large amount of data in the form of a galaxy catalog with verified redshift using slitless gratings. CSST is planned to start its observation around 2024, while MeerKAT will still be on full-time job before SKA which will begin full operations in 2028, and these two surveys will have a large overlapping survey area. Thus we believe MeerKAT H\textsc{i} intensity mapping and CSST galaxy survey would make promising cross-correlation detection in the coming future.

This paper is organized as follows:
in Section \ref{sec:simulation}, we introduce our method of creating mock data of MeerKAT H\textsc{i} intensity mapping and CSST spectroscopic galaxy surveys; in Section \ref{foreground removal}, we apply PCA algorithm to remove the foreground in H\textsc{i} intensity map; in Section \ref{power spectrum}, we calculate the galaxy auto and H\textsc{i}-galaxy cross power spectra, and discuss the signal compensation method for cross power spectrum; in Section \ref{cosmological constraint} we forecast the constraints on relevant cosmological parameters; we conclude our work and provide discussion in Section \ref{conclusion and discussion}.

\section{Mock data} \label{sec:simulation}
We generate MeerKAT intensity maps and CSST spectroscopic galaxy survey data using MultiDark cosmological simulations \citep{MultiDark_simulations}. MultiDark is a suite of $N$-body cosmological simulations which have been carried out by L-GADGET-2 code. 
Most simulations of this suite have $3840^3$ particles, with box sizes ranging from $250$ $\mathrm{Mpc}/h$ to $2500$ $\mathrm{Mpc}/h$. Based on the survey area and redshift of the MeerKAT observation plan, the Small MultiDark Planck simulation (SMDPL) has been chosen in this work. The box size of SMDPL is $400$ $\mathrm{Mpc}/h$, and halos in SMDPL boxes are identified through the halo finding code Friends-of-Friends (FOF) with relative linking length of $0.2$ \citep{fof}. The relevant simulation and cosmological parameters that SMDPL adopted are listed in Table~\ref{simulation paramters}, and its halo catalog can be acquired from the CosmoSim database\footnote{The data is available at \href{https://www.cosmosim.org/}{https://www.cosmosim.org/} }. 

In our work, we focus on the cosmology at $z=0.5$, which is one of the main observational target redshifts for both CSST and the MeerKAT $L$-band. So our mock data are generated from the snapshot70 of SMDPL, whose redshift $z\approx 0.5$. We also find that the $400$ $\mathrm{Mpc}/h$ box size of the snapshot70 corresponds to a survey of $\sim297\,\mathrm{deg}^{2}$ at $z=0.5$. Note that the non-flat sky effect may need to be considered for a $\sim300$ deg$^2$ sky coverage, but for simplicity, we still use the flat sky approximation in our mock data analysis. 

\begin{table}
    \centering
    \begin{tabular}{|c|c|}
        \hline
        Parameters & Values \\ \hline
        $L_{\mathrm{box}}$ & $400~\mathrm{Mpc}/h$ \\
        $N_p$ & $3840^3$\\
        $M_p$ & $9.63\times10^7\mathrm{M_{sun}}/h$\\
        $\epsilon$ & $1.5 \, \mathrm{kpc}/h$\\
        $h$ & $0.6777$\\
        $\Omega_{M}$ & $0.307$\\
        $\Omega_{B}$ & $0.048$\\
        $\Omega_{\Lambda}$ & $0.693$\\
        $n_s$ & $0.96$\\
        $\sigma_{8}$ & $0.8228$\\
        \hline
    \end{tabular}
    \caption{Simulation parameters for SMDPL.}
    \label{simulation paramters}
\end{table}

\subsection{H\textsc{i} intensity mapping with MeerKAT}

\begin{figure}
    \centering
    \includegraphics[width = 0.5\textwidth]{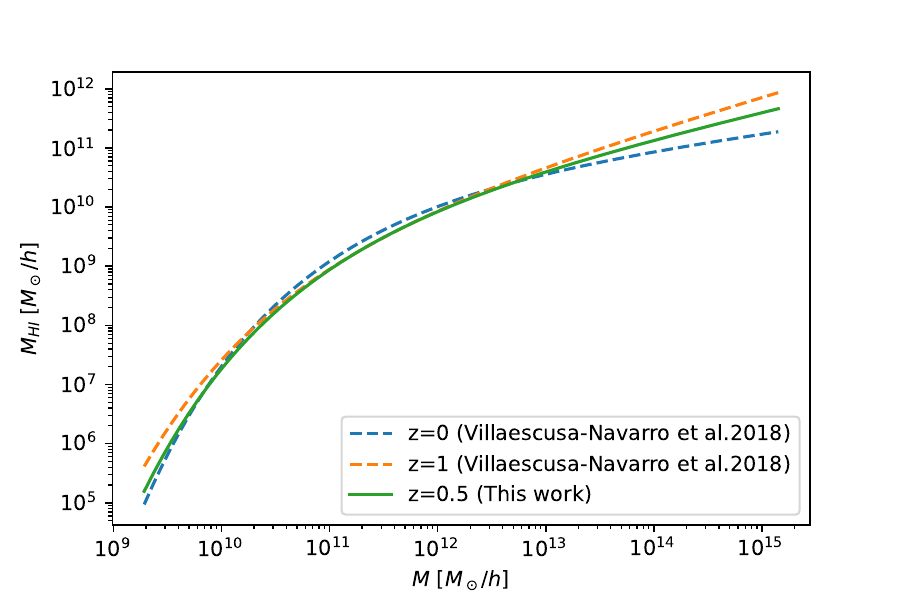}
    \caption{The H\textsc{i} mass $M_{\rm H\textsc{i}}$ and halo mass $M$ relation we use at $z=0.5$ (green curve), which is derived from the relations at $z=0$ (blue dashed curve) and $z=1$ (orange dashed curve) given in \citet{HI_calcution_parameters}.}
    \label{fig:HI-halo_mass_relation}
\end{figure}

Since H\textsc{i} could only survive from ultraviolet (UV) radiation in dense clumps in galaxies after the epoch of reionization, we assume that H\textsc{i} can only exist in halos hosting galaxies at $z=0.5$. We place the H\textsc{i} mass in the center of a halo, as has been proven to be reasonable in previous studies  \citep[see e.g.][]{HI_calcution_parameters}. Under this assumption, we construct a catalogue applying the halo H\textsc{i} mass function given by \citet{HI_calcution_parameters}, and it takes the form
\begin{equation}
     M_{\rm H\textsc{i}}(M,z)=M_0\left(\frac{M}{M_{\rm min}}\right)^{\alpha}\mathrm{exp}\left[-\left(\frac{M}{M_{\rm min}}\right)^{0.35}\right].
\end{equation}
Here $M$ is the halo mass, and we have three free parameters, i.e. $\alpha$, $M_{0}$ and $M_{\rm min}$, which determine the shape of the fitting curve at different redshifts. In order to get the values of these three parameters at z=0.5, we perform interpolation on the fitting values at $z=0$ and 1 given in \citet{HI_calcution_parameters}.
Then we find that $\alpha=0.42$, $M_{0}=2.50\times 10^{10} h^{-1}M_{\odot}$, $M_{\rm min}=1.13\times 10^{12} h^{-1}M_{\odot}$ at $z=0.5$. 
In Figure~\ref{fig:HI-halo_mass_relation}, we plot the $M_{\rm H\textsc{i}}-M$ relation at $z=0.5$ (green solid curve), and the relations at $z=0$ (blue dashed curve) and 1 (orange dashed curve) from \citet{HI_calcution_parameters} are also shown for comparison.

Then we can calculate the H\textsc{i} energy density parameter $\Omega_{\rm H\textsc{i}}$, which is expressed as 
\begin{equation}
    \Omega_{\rm H\textsc{i}}(z)=\frac{1}{\rho^{0}_{c}}\int n(M,z)M_{\rm H\textsc{i}}(M,z)dM,
\end{equation}
where $\rho_{c}^{0}$ is the critical density of the present Universe, and $n(M,z)$ is the halo mass function \citep{halo_mass_function}, which can be derived from our simulation. 
We find that $\Omega_{\rm H\textsc{i}}=6.73\times10^{-4}$ in our simulation, which agrees with the estimation of $\Omega_{\rm H\textsc{i}}-z$ relation given in literatures \citep[see e.g.][]{HI_calcution_parameters}.

\begin{figure*}
\begin{center}
\includegraphics[width=0.497\textwidth]{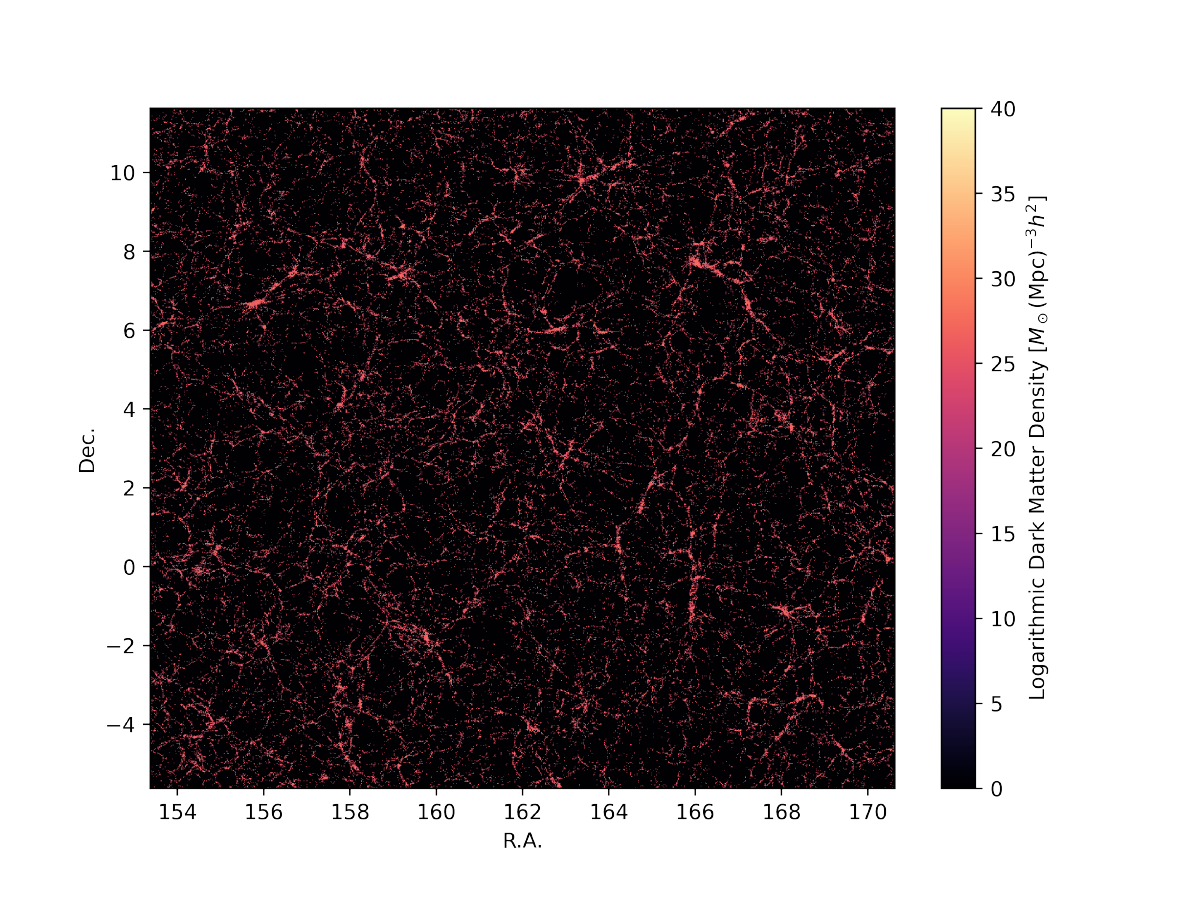}
\includegraphics[width=0.497\textwidth]{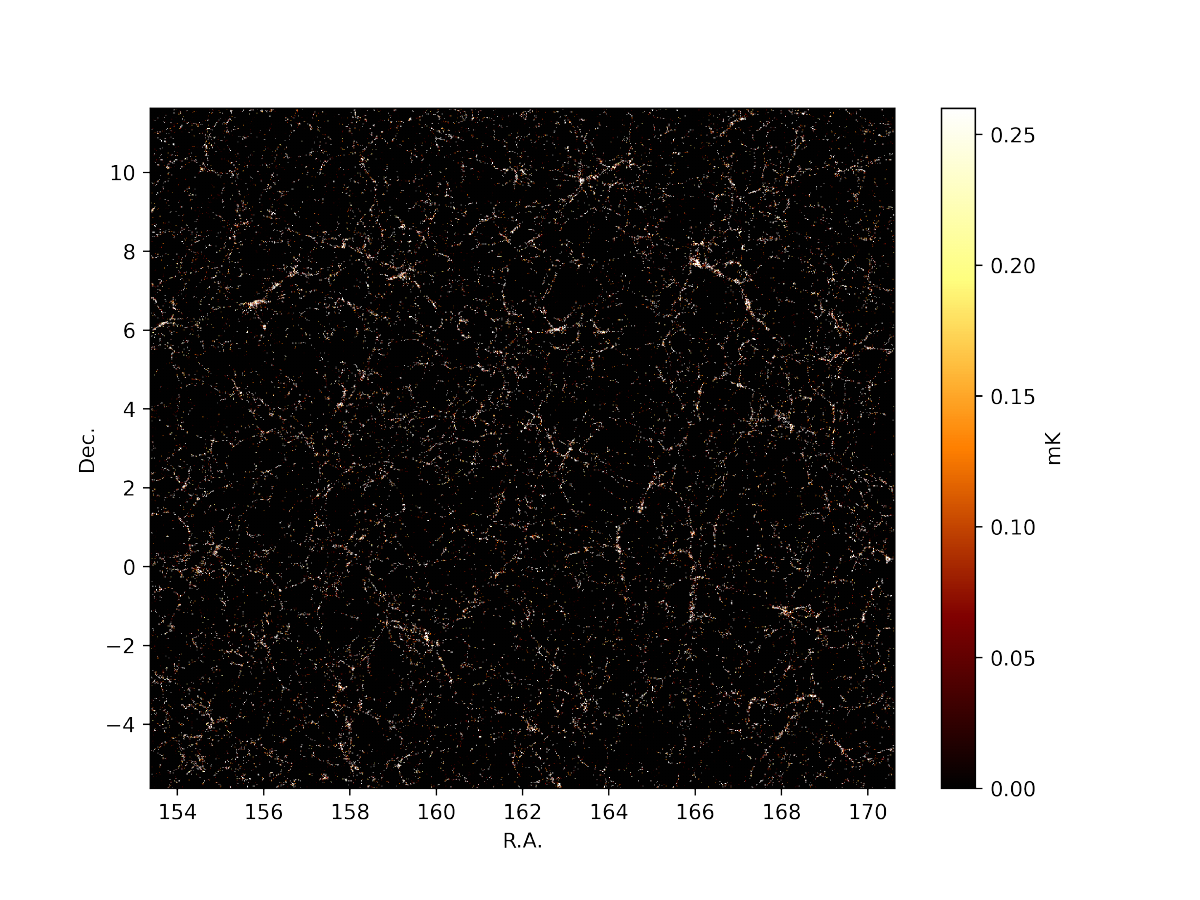}\\
\caption{{\it Left panel:} the dark matter distribution at $z=0.5$ in the simulation. {\it Right panel:} the corresponding map of H\textsc{i} brightness temperature. We can see that the H\textsc{i} map has similar structure as dark matter, and can be a tracer for the LSS.}
\label{fig:DM_distribution}
\end{center}
\end{figure*}

\begin{figure*}
\begin{center}
\includegraphics[width=0.49\textwidth]{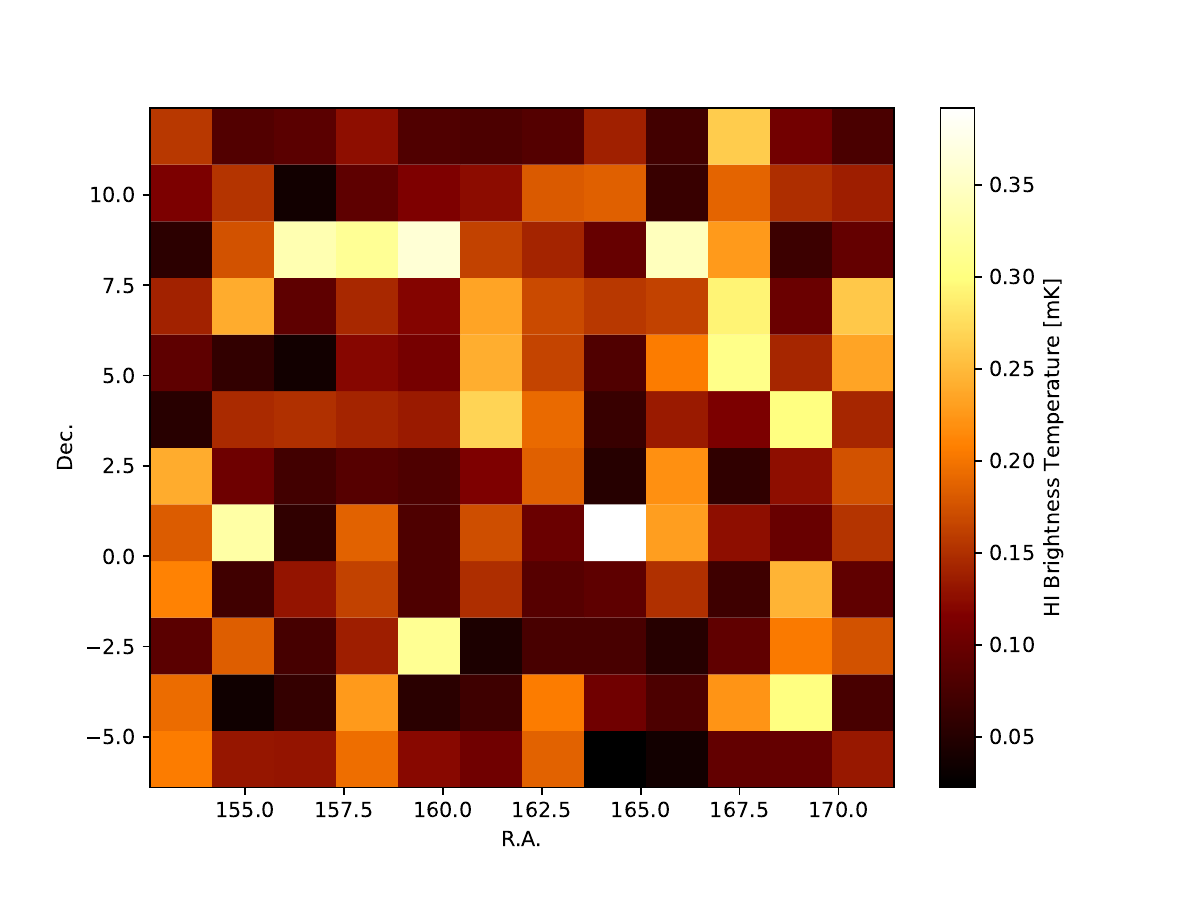}
\includegraphics[width=0.49\textwidth]{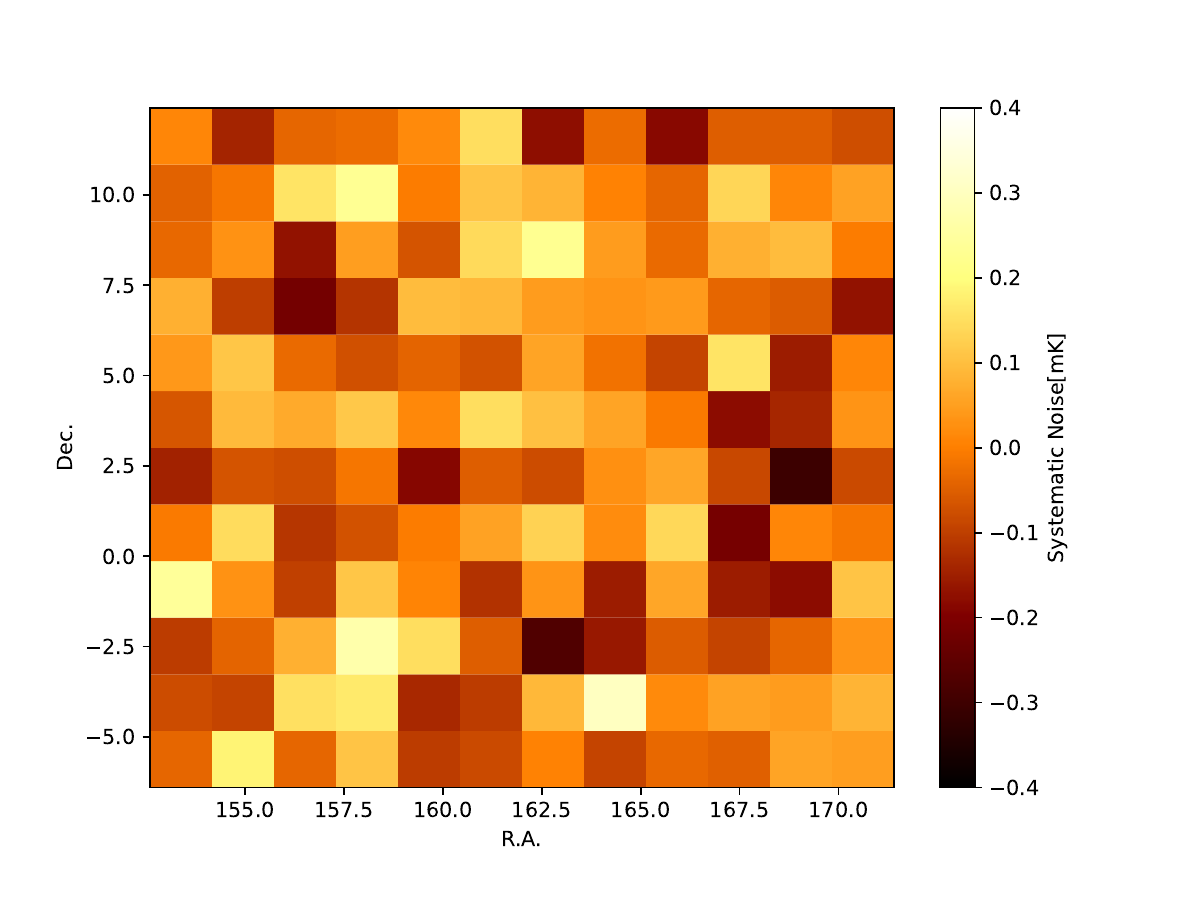}\\
\includegraphics[width=0.49\textwidth]{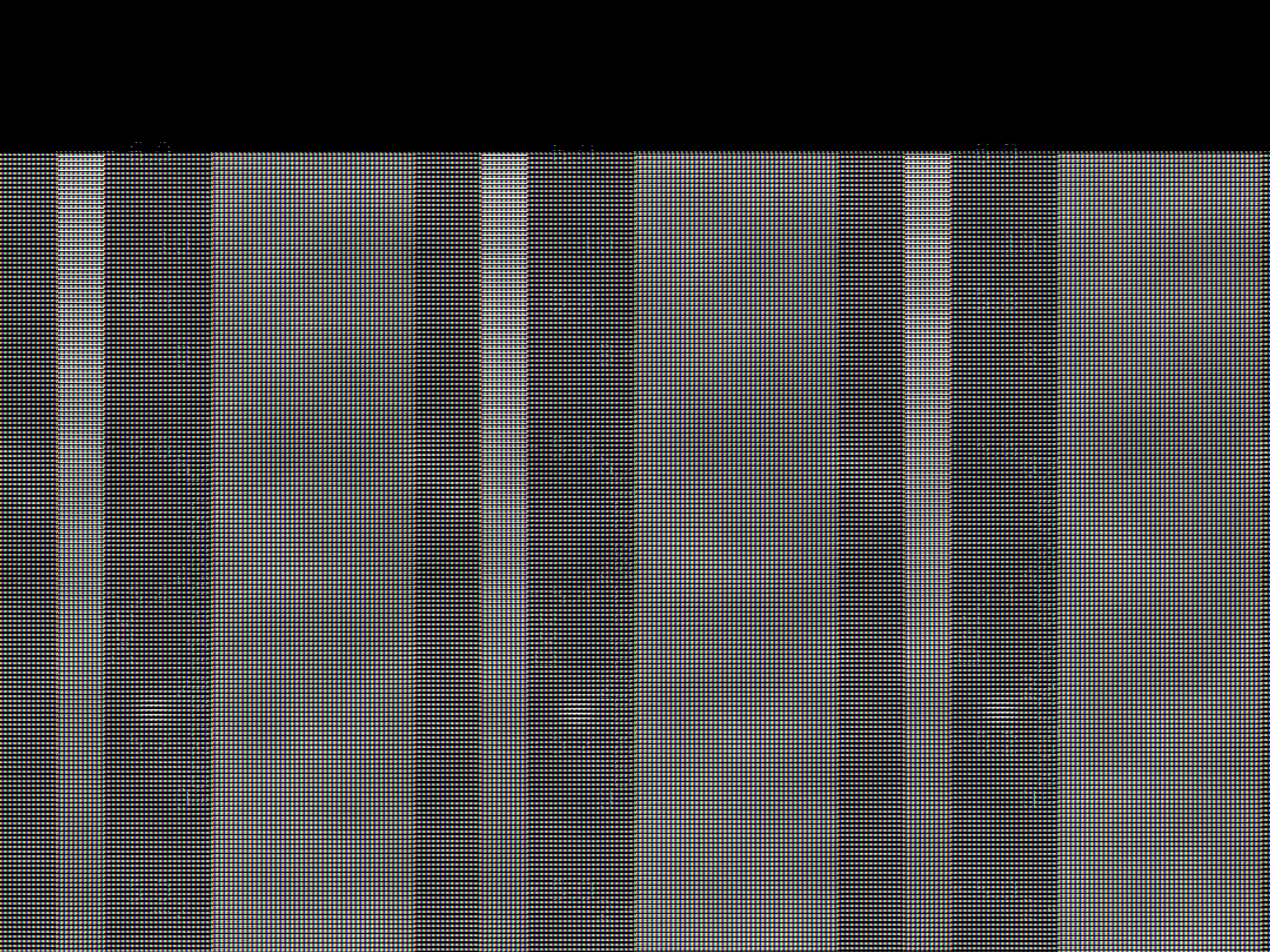}
\includegraphics[width=0.49\textwidth]{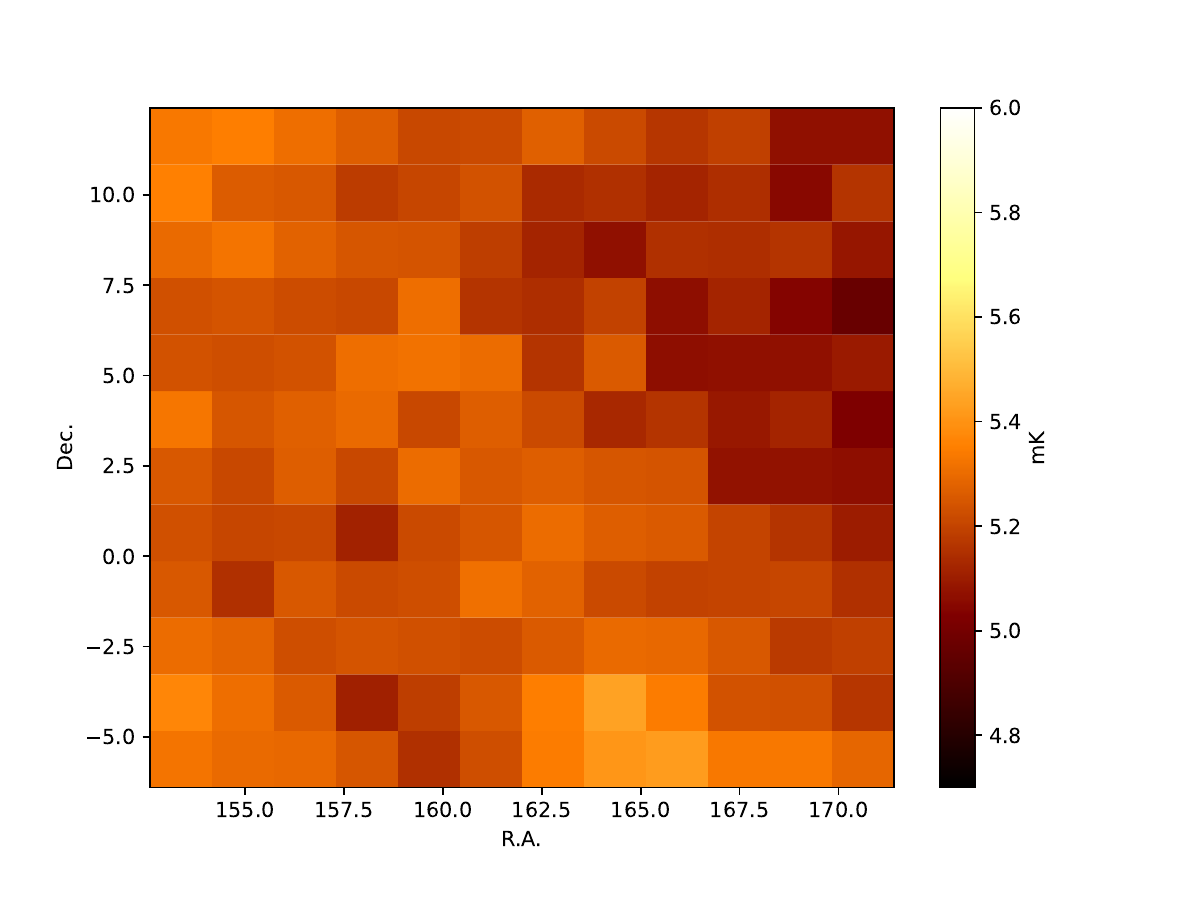}\\
\caption{The mock MeerKAT intensity maps for the central slice in the simulation at $\nu=946.7 ~\rm MHz$ or $z=0.5$. The upper left panel is the signal map of H\textsc{i} brightness temperature. The upper right panel features the map of Gaussian system noise. The bottom left panel is the total foreground map generated by GSM2016 at $\nu=946.7 ~\rm MHz$, including Galactic synchrotron emission, free-free emission, cold and warm dust thermal emission, the cosmic microwave background (CMB) anisotropy, and Galactic H\textsc{i} emission. The bottom right panel depicts the total sky map observed by MeerKAT containing all components we consider.}
\label{fig:4_figures}
\end{center}
\end{figure*}

Next, we can create the map of H\textsc{i} brightness temperature. The brightness temperature field $\delta_{T}$ traces the underlying matter fluctuations $\delta_{\rm m}$ as 
\begin{align}
    \delta_{T}(\bm{r},z)=\overline{T}_{\rm b}(z) b_{\rm H\textsc{i}}(z) \delta_{\rm m}(\bm{r},z)
    \label{delta_T}
\end{align}
where $\overline{T}_{\rm b}(z)$ is the mean H\textsc{i} brightness temperature at $z$, and $b_{\rm H\textsc{i}}$ is the H\textsc{i} bias, which can be estimated by
\begin{align}
    b_{\rm H\textsc{i}}(z)=\frac{\int n(M,z)b(M,z)M_{\rm H\textsc{i}}(M,z){\rm d} M}{\int n(M,z)M_{\rm H\textsc{i}}(M,z){\rm d}M}.
    \label{b_HI}
\end{align}
Here $b(M,z)$ is the halo bias. So for a voxel with position on the sky $\mathbf{r}$ and redshift $z$, its H\textsc{i} brightness temperature can be derived as
\begin{align}
    T_{\rm b}(\bm{r},z) & = 189\frac{h}{E(z)}\Omega_{\rm H\textsc{i}}(\bm{r},z)(1+z)^{2} \  [\mathrm{mK}] \nonumber \\
    & =T_0(z)\times \Omega_{\rm H\textsc{i}}(\bm{r},z) ,
\end{align}
where  $E(z)=H(z)/H_{0}$ represents the evolution of the Hubble parameter, and $T_0$ is a redshift dependent parameter which is defined as $T_0=189\frac{h}{E(z)}(1+z)^{2}$. Then $\overline{T}_{\rm b}(z)$ can be estimated by averaging $T_{\rm b}(\mathbf{r},z)$ at different positions in the simulation box. At $z=0.5$, we find that the corresponding mean H\textsc{i} brightness temperature is $\overline{T}_{\rm b}=0.145$ $\mathrm{mK}$ in our simulation. The brightness temperature of the H\textsc{i} distribution (right panel) and the corresponding dark matter distribution (left panel) in the simulation are shown in Figure~\ref{fig:DM_distribution}.

After obtaining the H\textsc{i} brightness temperature in the simulation box, our next step is to create the H\textsc{i} intensity maps with MeerKAT instrumental parameters and observational effects. Since the observable of H\textsc{i} intensity mapping is the H\textsc{i} brightness temperature of each voxel in the survey volume, we divide the survey volume (here this means our simulation box) into voxels that MeerKAT can observe. The details are as follows:
\begin{itemize}
    \item To divide frequency bins along the line of sight (LOS), we put the center of the box at $z=0.5$. As the box length of 400 $\mathrm{Mpc}/h$ is known, the redshift range of the survey volume can be calculated. We find that the redshift range of snapshot70 is 0.415 $\sim$ 0.590, corresponding to the observed H\textsc{i} frequency of 1004.14 MHz $\sim$ 893.30 MHz. This frequency range can be observed by the MeerKAT $L$-band with frequency resolution of 0.2 MHz, and it allows us to divide the survey volume into 554 bins, that each bin width is about 0.72 $\mathrm{Mpc}/h$. For simplicity, we assume that there is no redshift evolution in this range.
    \item As for the pixels perpendicular to LOS, since we plan to use the single-dish mode observation, resolution $\theta$ is defined by the full width at half maximum (FWHM) of the beam of an individual dish. Then the beam size or spatial resolution is given by
    \begin{align}
        \theta_{\rm b}=1.02\frac{\lambda_{\rm obs}}{D_{\rm dish}},
    \end{align}
where $\lambda_{\rm obs}$ is the observed wavelength, and $D_{\rm dish}$ is the dish aperture diameter. We find that the spatial resolution of MeerKAT at $z=0.5$ is 1.36 deg. Since the size of a simulation box is $400\times 400$ $(\mathrm{Mpc}/h)^{2}$ corresponding to a 297 $\mathrm{deg}^2$ survey area, the number of pixels in an H\textsc{i} map is found to be $12\times 12$ for MeerKAT single-dish mode observation. We note that the current spatial resolution given by the FWHM of the beam is a choice of simplicity, and more realistic resolution will be considered in the future work. Besides, since the beam size actually changes with frequency, it can introduce more complexity and challenges into the foreground subtraction. However, because our simulation snapshot has no redshift evolution, for simplicity, we do not consider the frequency dependence of the beam size, and set the pixel size of all the maps to be the same.
\end{itemize}

The H\textsc{i} signal intensity map obtained by MeerKAT at $z=0.5$ is displayed in the upper left panel of Figure~\ref{fig:4_figures}. In real observation, the H\textsc{i} intensity mapping will be contaminated by different components, such as system thermal noise, foreground emission from the Milky Way, radio frequency interference (RFI), etc., which can lower the SNR. Here we model the system thermal noise of a single-dish as Gaussian noise. Its root mean square (rms) noise temperature can be calculated as \citep{noise_calculation}
\begin{align}
    \sigma_{\rm T}=\frac{T_{\rm sys}}{\sqrt{\delta\nu t_{\rm tot}}}\frac{\lambda^2}{\theta_{\rm b}^2 A_{e}}\sqrt{A_{\rm S}/\theta_{\rm b}^2},
\end{align}
where $\delta \nu$ is the frequency interval, $t_{\rm tot}$ is the total observation time of the survey, $A_{e}$ is the effective collecting area of a dish, $A_{\rm S}$ is the survey area and $T_{\rm sys}$ is the system temperature which is usually described as a combination of four components, yielding
\begin{align}
    T_{\rm sys} = T_{\rm sky}(\nu)+T_{\rm spill}+T_{\rm atm}+T_{\rm rec}.
\end{align}
The mean sky temperature can be approximated by $T_{\rm sky} = 2.725+1.6(\nu/\mathrm{GHz})^{-2.75}$, and $T_{\rm spill}$, $T_{\rm atm}$ and $T_{\rm rec}$ represent spillover temperature, atmosphere temperature and receiver temperature, respectively. The values of these parameters we adopt are listed in Table.~\ref{MeerKAT specifications}, and then we obtain $\sigma_{\rm T}=0.102\, \rm mK$ at $\nu = 946.7\, \rm MHz$ (z=0.5). The corresponding map of Gaussian system noise is shown in the upper right panel of Figure~\ref{fig:4_figures}.


\begin{table}
    \centering
    \begin{tabular}{lc}
        \hline
        Parameters & Values \\ \hline
        Antennas & All 64 MeerKAT dishes \\
        Observation mode & Single-dish\\
        Dish diameter & 13.5m \\
        System temperature & ~20 K\\
        $L$-band Frequency range & 856-1712 MHz\\
        Frequency resolution & 0.2MHz\\
        Survey time & 200 hr per dish\\
        \hline
    \end{tabular}
    \caption{Specifications of the MeerKAT observations}
    \label{MeerKAT specifications}
\end{table}

The foreground emission from the Milky Way is actually the main challenge to H\textsc{i} intensity mapping. The brightness temperature of foregrounds can be more than 4 orders of magnitude brighter than H\textsc{i} signal, so its effect has to be seriously taken into account in our forecast of MeerKAT observation. Here we generate the foreground emission using the GSM2016 model \citep{foreground_model}. GSM2016 is an improved model of the original GSM. It uses an extended PCA algorithm to identify different components in the diffuse Galactic emission. Six components of Galactic emission that match the known physical emission mechanisms are obtained, i.e. synchrotron emission, free-free emission, cold and warm dust thermal emission, the CMB anisotropy and Galactic H\textsc{i} emission. This algorithm allows it to make use of 29 sky maps from $10\,\rm MHz$ to $5\, \rm THz$, and make interpolation to get full-sky map of any frequency in this frequency range.

To apply the foreground model on our H\textsc{i} map, the coordinates of the survey area have to be set. According to the previous work \citep{Meerkat_temperature}, we set our survey area at $153.38^\circ <$ R.A. $<170.62^\circ$ and $-5.62^\circ < $Dec.$ < 11.62^\circ$, mostly intersecting with the WiggleZ 11hr field \citep{WiggleZ2010} \citep{WiggleZ2018}. Note that this choice of survey area is only for discussion here, since this area has relatively low Galactic emission in the full-sky map. For future observation of MeerKAT-CSST cross-correlation, the target survey area can be chosen from anywhere in the overlapping region of the MeerKAT and CSST survey area with relatively low foreground emission. We generate foreground maps at each frequency bin, and then interpolate them to the center of each voxel. 

The foreground map for the survey area at $\nu=946.7\, \rm MHz$ ($z=0.5$) is shown in the lower left panel of Figure~\ref{fig:4_figures}. We find that the foreground contamination we consider is about 4 orders of magnitude brighter than the H\textsc{i} signal. 
After combining the H\textsc{i} signal, foreground emission and system noise maps, we obtain the total sky map observed by MeerKAT. The mock total observational map is displayed in the lower right panel of Figure~\ref{fig:4_figures}. Since the H\textsc{i} signal has totally drowned in the contamination, the foreground contaminant subtraction algorithms have to be applied to extract the H\textsc{i} signal. We will discuss the foreground removal method in the next section.

\subsection{Galaxy survey with CSST}

\begin{figure}
    \centering
    \includegraphics[width = 0.5\textwidth]{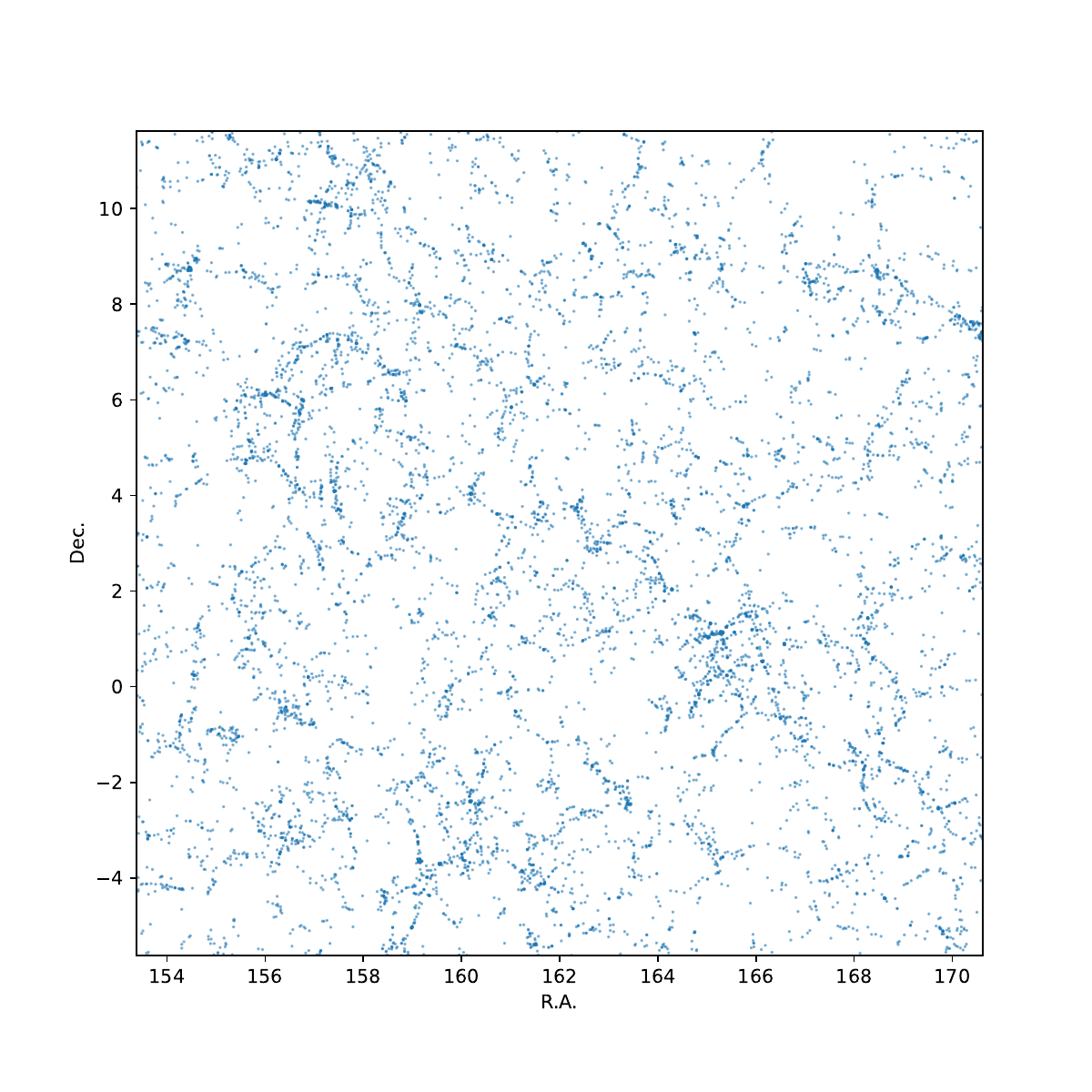}
    \caption{The mock galaxy map observed by CSST spectroscopic survey at $z=0.5$, which is in the same survey area as the MeerKAT H\textsc{i} intensity mapping survey.}
    \label{fig:galaxy_distribution}
\end{figure}

We use the same simulation data SMDPL snapshot70 to create the mock data of CSST spectroscopic galaxy survey. We utilize Python package Halotools\footnote{\href{https://halotools.readthedocs.io/}{https://halotools.readthedocs.io/} }
to generate a galaxy distribution for each dark matter halo in the simulation. First, the structure of a cold dark matter (CDM) halo can be described by an NFW profile \citep{NFWprofile}. The halo concentration-mass relation under the NFW profile is fitted by \citep{halo_con}
\begin{align}
    \mathrm{log_{10}} c_{\rm vir}=a+b\mathrm{log_{10}}(M_{\rm vir}/[10^{12}h^{-1}M_{\odot}]),
\end{align}
where $M_{\rm vir}$ is the halo virial mass, $c_{\rm vir}$ is the concentration of the corresponding halo, and $a$ and $b$ are the fitting parameters, which are expressed as
\begin{align}
    a=0.537+(1.205-0.537)\mathrm{exp}(-0.718z^{1.08}),
\end{align}
\begin{align}
    b=-0.097+0.024z.
\end{align}
After obtaining the halo concentration, the halo occupation distribution (HOD) model can be applied to get galaxy distribution. The HOD model can determine the population of central galaxy and satellite galaxies in a given halo. The central galaxy occupation statistics is given by \citep{hod}
\begin{align}
    \langle N_{\rm cen} \rangle=\frac{1}{2}\left(1+\mathrm{erf}\left( \frac{\mathrm{log_{10}}M-\mathrm{log_{10}}M_{\rm min}}{\sigma_{\mathrm{log_{10}}M}}\right) \right),
    \label{eq:N_cen}
\end{align}
and central galaxies are assumed to reside at the centers of the host halos. On the other hand, the distribution of satellite galaxies is written as
\begin{align}
    \langle N_{\rm sat} \rangle = \left( \frac{M-M_0}{M_1}\right)^{\alpha}. 
    \label{eq:N_sat}
\end{align}
When redshift, cosmological model and the threshold of galaxy absolute magnitude are set, the values of the parameters in Eq.~\eqref{eq:N_cen} and \eqref{eq:N_sat} are calculated by the Halotools package based on the model published in \cite{hod}. At $z=0.5$, we set the threshold of galaxy absolute magnitude to be $-19.5$, and then these parameters are expressed as
$\mathrm{log_{10}}M_{\rm min}=11.35$, 
$\sigma_{\mathrm{log_{10}}M}=0.28$, 
$\mathrm{log_{10}}M_{0}=11.69$, 
$\mathrm{log_{10}}M_{1}=13.01$ and $\alpha=1.06$.

The final step in generating a galaxy catalog is to determine which galaxies can be observed by CSST. We assign luminosity to galaxies using the relation between host halo mass and galaxy luminosity, which is given by \citep{mass_luminosity_relation}

\begin{align}
    L_{\rm group}=L_{A}\left(\frac{M}{\mathrm{M_{\odot}}}\right)^{0.88}h^{-2}\ \mathrm{L_{\odot}},
\end{align}
\begin{align}
    L_{\rm cen}=L_{0}\frac{(M/M')^{a}}{[1+(M/M')^{bc}]^{1/c}} h^{-2}\ \mathrm{L_{\odot}},
\end{align}
and for simplicity, if assuming all satellite galaxies have the same luminosity, we have
\begin{align}
    L_{\rm sat}=\frac{1}{N_{\rm sat}}(L_{\rm group}-L_{\rm cen}).
\end{align}
Here $L_{\rm group}$, $L_{\rm cen}$, and $L_{\rm sat}$ are the luminosities of a galaxy group, central galaxy, and satellite galaxy, respectively. $N_{\rm sat}$ is the number of satellite galaxies in a galaxy group. The parameter values are chosen to be 
$L_{A}=0.3~\mathrm{L_{\odot}}$,
$L_0=2.8\times 10^{9}~\mathrm{L_{\odot}}$ \citep{hod}, 
$M'=3.7\times 10^{9}~ h^{-1}\mathrm{M_{\odot}}$, 
$a=29.78$, 
$b=29.5$ and 
$c=0.0255$. \citep{mass_luminosity_relation}.
Then the galaxy luminosity can be converted to the magnitudes in CSST spectroscopic bands. Since the magnitude limit of the CSST spectroscopic survey is $\sim$23 mag \citep{CSS_OS,Zhan2021}, 
galaxies whose magnitude is under this limit can be selected to form the CSST spectroscopic galaxy survey catalog. After the selection, we find that the galaxy number density in the simulation box is $9.07\times 10^{-3} ~ ({\rm Mpc}/h)^{-3}$, which is in good agreement with the result in previous works \citep[e.g.][]{CSS_OS}. The mock map of the CSST spectroscopic galaxy survey at $z=0.5$ is depicted in Figure.~\ref{fig:galaxy_distribution}.

\section{foreground removal}\label{foreground removal}

\begin{figure}
    \centering
    \includegraphics[width = 0.5\textwidth]{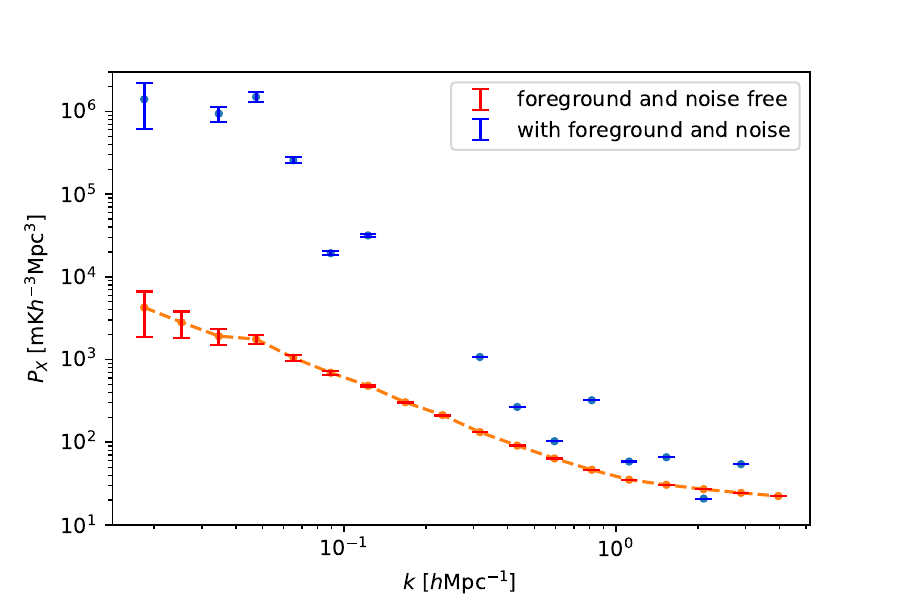}
    \caption{The cross-correlation power spectra of the MeerKAT H\textsc{i} raw (blue data points) and signal (red data points) intensity maps with the corresponding CSST spectroscopic galaxy map. The missing data points in the raw cross power spectrum (blue data points) have negative or small values, so they are not shown in the figure.}
    \label{fig:cross-correlation_about_foreground}
\end{figure}

The signal extraction of H\textsc{i} intensity mapping highly relies on foreground removal efficiency. Theoretically, cross-correlation with other tracers (e.g. galaxies and other emission lines) could be a good way to extract the H\textsc{i} signal and reduce the effects of foregrounds and system noise, since the foregrounds and instrumental noise of different wave bands in different surveys should be uncorrelated. However, it is found that it will be problematic if directly cross-correlating the raw intensity map with other surveys. In Figure~\ref{fig:cross-correlation_about_foreground}, we show the results of our mock MeerKAT H\textsc{i} raw (blue data points) and signal (red data points) maps cross-correlated with the mock CSST spectrographic galaxy map.
We can see that the effect of foreground contamination is still too huge to extract correct cosmological information, as there is large deviation at all scales between the two curves. This indicates that extra foreground removal methods should be performed before cross-correlation. 

Many methods of foreground removal have been discussed in previous works. These include 
blind foreground subtraction algorithms, such as
PCA and Singular Value Decomposition (SVD)\citep{PCA1,PCA2,PCA-SVD,SVD}, 
ICA\citep{ICA}, 
correlated component analysis (CCA)\citep{CCA}, 
extended ICA \citep{extend_ICA} and
FASTICA \citep{FASTICA}, 
non-parametric Bayesian methods, like Gaussian Progress Regression (GPR) \citep{GPR1,GPR2},
and methods assuming some physical properties of the foregrounds, such as polynomial/parametric-fitting \citep{fitting2,fitting1}.
Here we use the PCA/SVD algorithm to perform foreground removal. Since the PCA method is based on identifying different correlations of corresponding components in the frequency domain, in principle, this method can distinguish the foregrounds from the H\textsc{i} signal by their different frequency smoothness. Besides, PCA does not require much knowledge about the models of data components, which is suitable in our case. On the other hand, SVD is a similar method that can be applied on a data matrix, which can obtain similar results as PCA but with fewer calculation steps.

To apply our foreground removal procedure, first we transform the simulation result into data matrix $X$ with dimensions $N_{\nu}\times N_{\rm p}$. Here $N_{\nu}$ is the number of frequency channels, and $N_{\rm p}$ is the number of pixels in a frequency channel of the intensity map. 
Then SVD can decompose the data matrix $X$ in the form 
\begin{equation}
    X=W^{\mathrm{T}}\Sigma R,
    \label{eq:SVD_decompose}
\end{equation}
where $W^{\mathrm{T}}$ and $R$ are called left and right singular vectors, respectively, and $\Sigma$ is a rectangular diagonal matrix of singular values. $W^{\mathrm{T}}$ and $R$ are unitary matrices, which are defined as $WW^{*}=1$ and $RR^{*}=1$, where asterisk denotes conjugate transpose.
Generally, when dealing with a complex valued matrix $X$, Equation~\eqref{eq:SVD_decompose} takes the form of $X=W^{\mathrm{*}}\Sigma R$. But since our data matrix $X$ is real, we use transposed matrix $W^{\mathrm{T}}$ to substitute $W^{\mathrm{*}}$.

Singular vectors $W^{\mathrm{T}}$ and singular values $\Sigma$ are equivalent to the eigenvectors and eigenvalues in PCA, respectively. So, we rank the singular vectors in decreasing order of their corresponding singular values to identify the principal components of the data matrix, i.e. the foregrounds. 

\begin{figure}
    \centering
    \includegraphics[width = 0.5\textwidth]{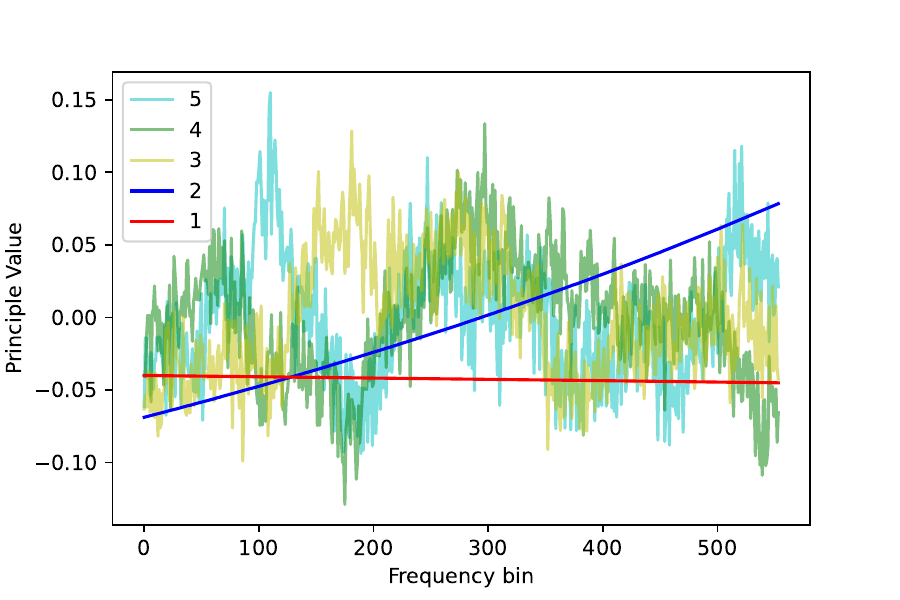}
    \caption{The first five principal components of data matrix $X$ decomposed by the PCA/SVD method.}
    \label{fig:Principle_components}
\end{figure}

\begin{figure*}
\begin{center}
\includegraphics[width=0.497\textwidth]{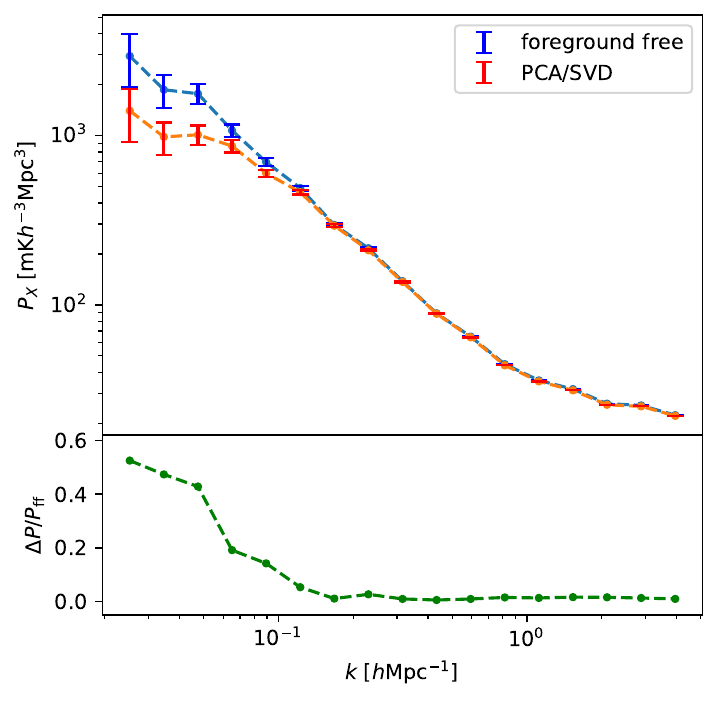}
\includegraphics[width=0.497\textwidth]{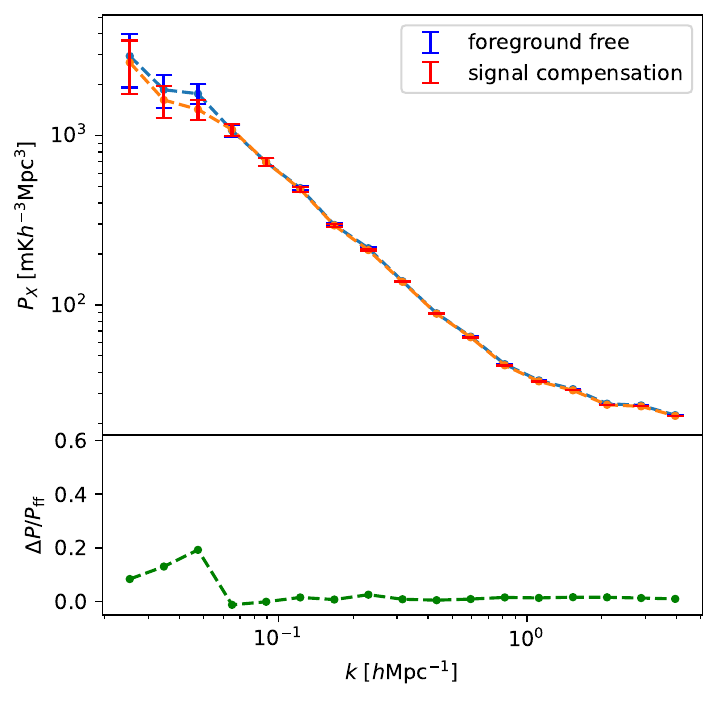}\\
\caption{{\it Left panel:} the MeerKAT-CSST cross power spectra of H\textsc{i} foreground free (blue data points) and foreground removal by PCA (red data points). {\it Right Panel:} the same as the left panel but considering signal compensation after PCA foreground removal (red data points). The corresponding relative errors are also shown in green dashed curves in the lower panels, where $\Delta P$ is the difference between the two power spectra in the upper panels.}
\label{fig:PCA_vs_compensation}
\end{center}
\end{figure*}

Then we compose an $N_{\nu}\times m$ projection matrix $W'$ with the first $m$ columns of $W$, where $m$ is the number of components which are thought to be foregrounds. In Figure~\ref{fig:Principle_components}, we show the first five principal components decomposed from the data matrix $X$. We notice that the first two components are relatively smooth in frequency smoothness, so they are identified as foreground, i.e. $m=2$.
The dominant principal components will be obtained when the data matrix $X$ is projected onto the projection matrix $W'$ by
\begin{align}
    U = W'^{\mathrm{T}}\cdot X,
\end{align}
\begin{align}
    V=W'\cdot U.
\end{align}
Here $U$ is the foreground information constructed from the data matrix. Then the H\textsc{i} signal can be recovered as
\begin{align}
    S_{\rm H\textsc{i}}=X-V.
\end{align}
At last, the recovered signal is projected back to the original map position for obtaining the foreground-removed map. 
In principle, the foreground-removed map is composed of H\textsc{i} signal and system noise. The effect of the PCA procedure can be indicated more clearly in a line intensity power spectrum as we discuss in the next section.

\section{power spectrum}\label{power spectrum}
Here we introduce the process of line intensity power spectrum estimation. We consider the cross-correlation using the method based on \citet{power_spectrum_equation}. The galaxy survey and intensity mapping data are converted into galaxy over-density and brightness over-temperature contrasts respectively by
\begin{align}
    \delta_{g}(\bm{x}_{i})=\frac{N(\bm{x}_{i})-\langle N \rangle}{\langle N \rangle},
\end{align}
\begin{align}
    \delta_{T}(\bm{x}_{i})=T_{\mathrm{H\textsc{i}}}(\bm{x}_{i})-\langle T_{\mathrm{H\textsc{i}}} \rangle,
\end{align}
where the angled brackets denote mean values. 
The Fast Fourier Transforms of $N(\bm{x}_{i})$ and $T_{\mathrm{H\textsc{i}}}(\bm{x}_{i})$ are given by
\begin{align}
    \Tilde{N}(\bm{k})= \sum_i N(\bm{x}_i) e^{i\bm{k}\cdot \bm{x}_i},
\end{align}
\begin{align}
    \Tilde{T}(\bm{k}) = \sum_i T_{\rm H\textsc{i}}(\bm{x}_{i})e^{i\bm{k}\cdot \bm{x}_i}.
\end{align}
Then our estimator for the galaxy auto-correlation power spectrum $P_{g}$ and the cross power spectrum between the gridded galaxy distribution and intensity map $P_{\times}$ at wavevector $\bm{k}$  are 

\begin{align}
    P_{g}(\bm{k})=V\langle \delta_{g}(\bm{k})\delta_{g}^{*}(\bm{k})\rangle-P_{\rm SN},
\end{align}
\begin{align}
    P_{\times}(\bm{k})= V\, \mathrm{Re}\{\delta_{g}(\bm{k}) \delta_{T}^{*}(\bm{k})\}.
\end{align}
Here $V=400^{3} (\mathrm{Mpc}h^{-1})^{3}$ is the survey volume and $P_{\rm SN}$ is the shot noise term for the galaxy survey, which can be estimated by $P_{\rm SN}=1/\langle N \rangle$. The error for the corresponding power spectrum is given by \citep{error_1994, power_spectrum_equation}
\begin{align}
    \sigma_{P_g}(k)=\frac{2\pi}{\sqrt{Vk^{2}\Delta k}}(P_{g}(k)+P_{\rm SN}),
\end{align}
\begin{align}
    \sigma_{P_{\times}} = &\frac{2\pi}{\sqrt{2Vk^{2}\Delta k}}\times \sqrt{P_{\times}^{2}(k)+(P_{T}(k)+P_{\rm N}(k))(P_{g}(k)+P_{\rm SN})},
\end{align}
where $\Delta k$ is k-bin width, $P_T$ is the H\textsc{i} brightness temperature power spectrum and $P_{\rm N}(k)$ is the power of system noise.

After performing the PCA/SVD foreground subtraction, we estimate and show the cross power spectra of CSST galaxy with foreground-free (blue data points) and foreground-subtracted (red data points) maps in MeerKAT H\textsc{i} intensity mapping survey, in the left panel of Figure~\ref{fig:PCA_vs_compensation}. We can find that signal loss can be caused by the PCA procedure, and it becomes severe especially at large scales that we are interested in. Therefore, the over-eliminated signal must be compensated.

\begin{figure}
    \centering
    \includegraphics[width = 0.5\textwidth]{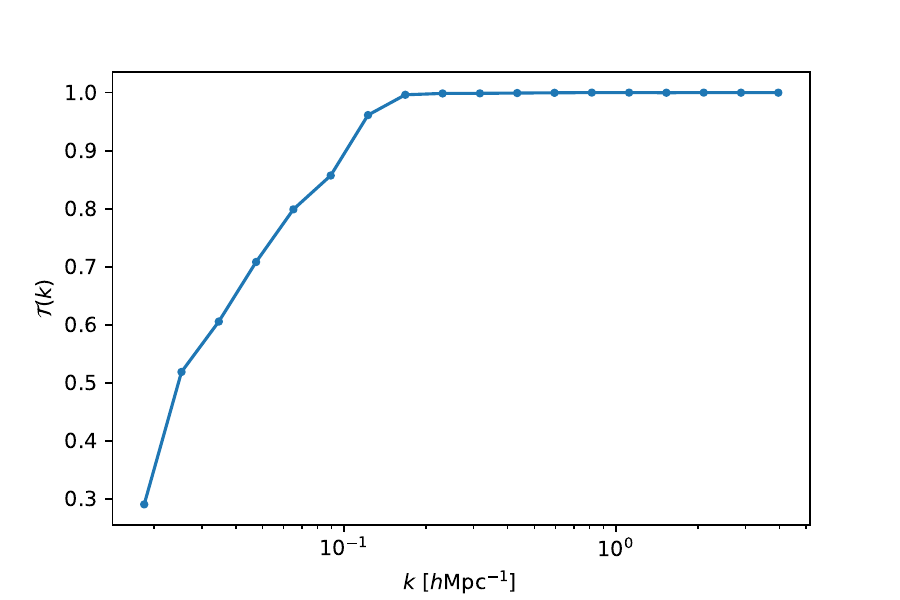}
    \caption{The transfer function $\mathcal{T}(k)$ estimated by 100 H\textsc{i} intensity mapping and corresponding galaxy survey mock data to compensate the signal loss after PCA foreground removal.}
    \label{fig:transfer_function}
\end{figure}

We compensate the cross power spectrum based on the method given in \citep{MeerKAT&WiggleZ}, and the procedure can be described as followS:
\begin{itemize}
    \item First, we generate mock data of halos. The mock halo catalogs include the information on mass and position, which follows the same matter power spectrum and halo mass function as SMDPL simulation.
    \item After that we calculate the H\textsc{i} brightness temperature of mock data using the H\textsc{i} model and MeerKAT observational effect. So, the mock H\textsc{i} intensity map data are obtained and further transformed into mock data matrix $Y$ with the same dimensions of data matrix $X$.
    \item Then the mock data matrix $Y$ is injected into the data matrix $X$. We apply the PCA clean on this data combination with the same projection matrix $W'$ we used in previous PCA. Then the foreground-removed mock data can be written as
    \begin{align}
        Y_{\rm c}=[Y+X ]_{\mathrm{PCA}} - S_{\rm H\textsc{i}}.
    \end{align}
    So, we can determine the signal loss of the cross power spectrum between $Y$ and $Y_{\rm c}$.
    \item Finally, the transfer function is constructed as
    \begin{align}
        \mathcal{T}(k) = \frac{\mathcal{P}(Y_{\rm c},Y_{g})}{\mathcal{P}(Y,Y_{g})},
    \end{align}
    where $\mathcal{P}()$ denotes the cross power spectrum and $Y_g$ is the corresponding mock galaxy data.
\end{itemize}
To compensate the signal loss, we construct the transfer function $\mathcal{T}(k)$ by generating 100 H\textsc{i} intensity mapping mock data and corresponding galaxy survey data. The result of the transfer function is shown in Figure~\ref{fig:transfer_function}.

The cross-correlation compensated by transfer function is displayed in the right panel of Figure.~\ref{fig:PCA_vs_compensation}. 
We can find that the signal compensation method we use is efficient, and that the compensated power spectrum is very consistent with the foreground-free power spectrum within $1\sigma$. Although over-compensation may happen due to large variance in the low $k$ range, the transfer function is reliable enough that the effect of signal loss can be effectively reduced. 

\section{cosmological constraint}\label{cosmological constraint}

\begin{figure}
    \centering
    \includegraphics[width = 0.5\textwidth]{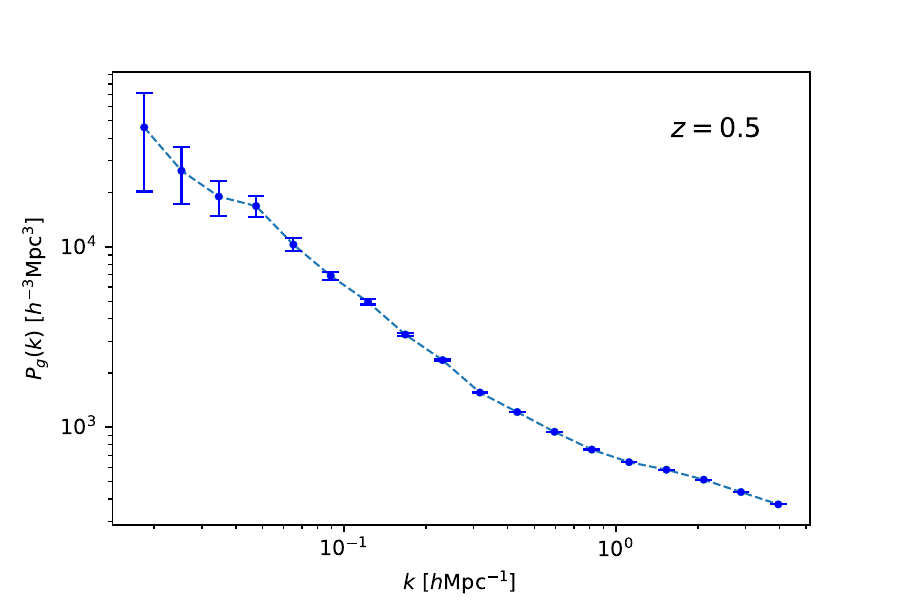}
    \caption{The galaxy auto power spectrum of CSST spectroscopic galaxy survey at $z=0.5$, which is derived from the simulation.}
    \label{fig:galaxy_power_spectrum}
\end{figure}

\begin{figure}
    \centering
    \includegraphics[width = 0.5\textwidth]{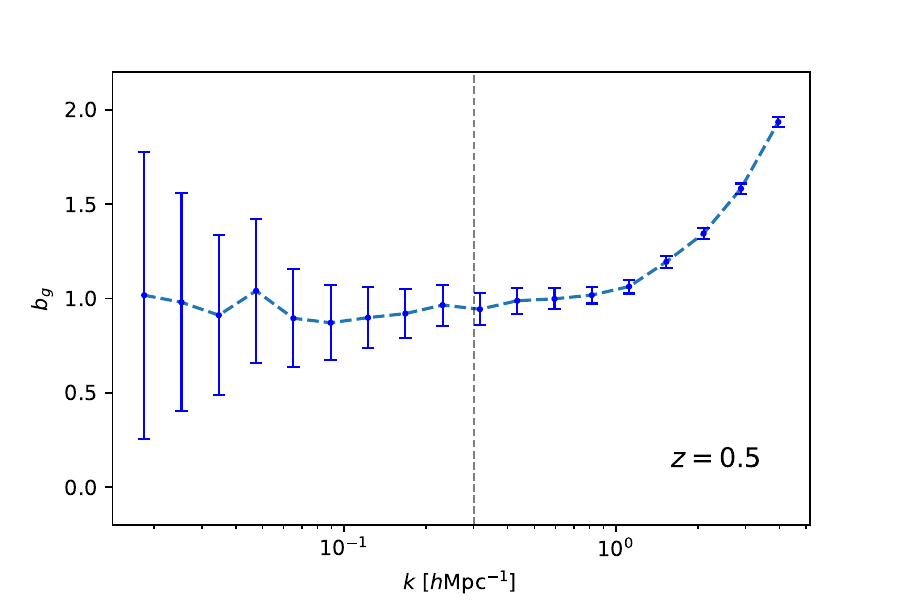}
    \caption{The galaxy bias factor $b_g$ for optical galaxies. The vertical dashed line is $k=0.3\, h^{-1}\mathrm{Mpc}$ that signifies the boundary of the linear scales we consider.}
    \label{fig:bg}
\end{figure}


\begin{table}
    \centering
    \begin{tabular}{|c|c|c|}
        \hline
        Parameters & Average Value & Error \\ \hline
        $b_{g}$& $0.9307$ & $\pm 0.0648$\\
        $10^{3}\Omega_{H\textsc{i}}b_{H\textsc{i}}b_{g}r_{H\textsc{i},g}$ & $0.4531$ & $\pm0.0045$ \\
        $10^{3}\Omega_{H\textsc{i}}b_{H\textsc{i}}r_{H\textsc{i},g}$ & $0.4812$ & $\pm0.0048$\\
        \hline
    \end{tabular}
    \caption{The value and error of $b_{g}$, $\Omega_{\rm H\textsc{i}}b_{\rm H\textsc{i}}b_{g}r_{{\rm H\textsc{i},}g}$ and $\Omega_{\rm H\textsc{i}}b_{\rm H\textsc{i}}r_{{\rm H\textsc{i},}g}$.}
    \label{cosmological parameters}
\end{table}

\begin{figure*}
\begin{center}
\includegraphics[width=0.497\textwidth]{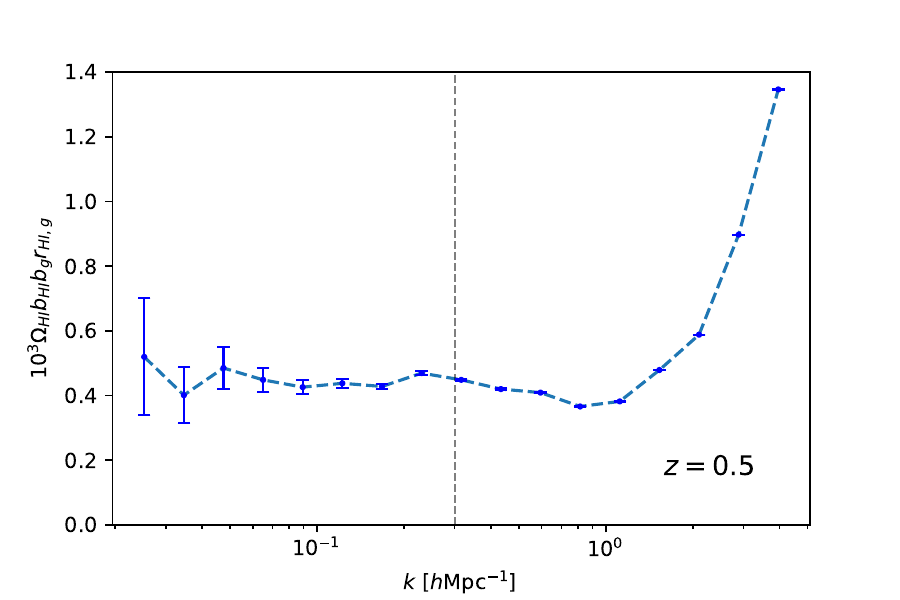}
\includegraphics[width=0.497\textwidth]{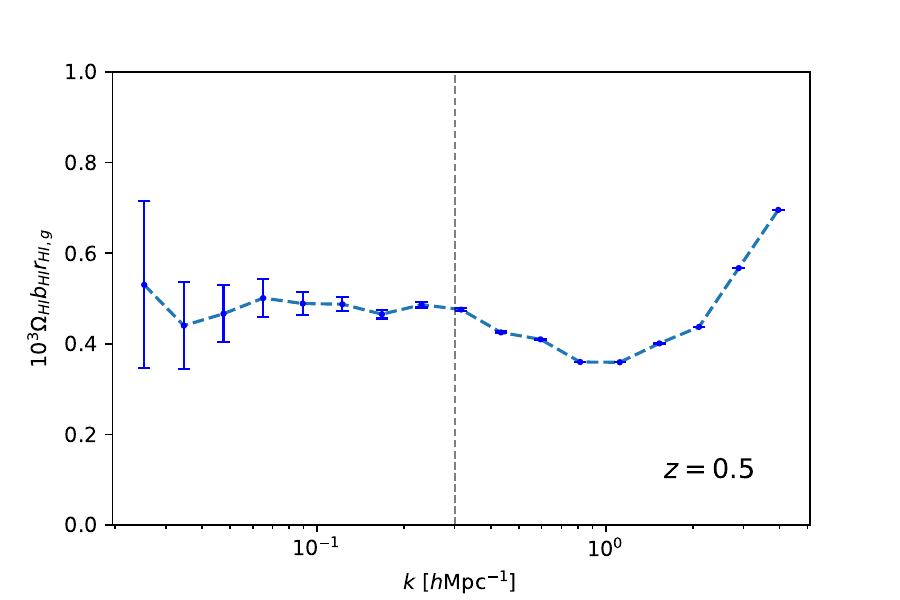}\\
\caption{The constraint results of $\Omega_{\rm H\textsc{i}}b_{\rm H\textsc{i}}b_{g}r_{{\rm H\textsc{i}},g}$ (left panel) and $\Omega_{\rm H\textsc{i}}b_{\rm H\textsc{i}}r_{{\rm H\textsc{i}},g}$ (right panel) from the cross-correlation of MeerKAT H\textsc{i} intensity mapping and CSST galaxy surveys. The vertical dashed lines are $k=0.3\, h^{-1}\mathrm{Mpc}$ that indicate the boundary of the linear regime we consider.}
\label{fig:Obbr_Obr}
\end{center}
\end{figure*}

After obtaining the cross power spectrum of MeerKAT H\textsc{i} intensity mapping and CSST spectroscopic galaxy surveys, we can explore the constraint power on cosmological parameters. Theoretically, the power spectra of different tracers have a similar relation to the matter power spectrum. The galaxy auto power spectrum $P_{g}(k)$ is related to the matter power spectrum $P_{\rm m}(k)$ as
\begin{align} \label{eq:Pg}
    P_g(k) = b_{g}^{2} P_{\rm m}(k),
\end{align}
where $b_{g}$ is the galaxy bias. On the other hand, the relation of the H\textsc{i} intensity auto power spectrum $P_{T}(k)$ and the matter power spectrum can be written as
\begin{align}
    P_{T}(k) &= \overline{T}_{b}^{2}b_{\rm H\textsc{i}}^{2}P_{\rm m}(k) \nonumber \\
    &= T_{0}^{2}\Omega_{\rm H\textsc{i}}^{2}b_{\rm H\textsc{i}}^{2}P_{\rm m}(k). 
\end{align}
Here $b_{\rm H\textsc{i}}$ is the H\textsc{i} bias, and $\overline{T}_{b}$ is the mean H\textsc{i} brightness temperature in the Universe.
Then the cross power spectrum $P_{\times}(k)$ is given by
\begin{align}
    P_{\times}(k)=T_{0}\Omega_{\rm H\textsc{i}}b_{\rm H\textsc{i}}b_{g}r_{{\rm H\textsc{i}},g}P_{\rm m}(k),
\end{align}
where $r_{{\rm H\textsc{i}},g}$ is the H\textsc{i}-galaxy correlation coefficient that indicates the correlation strength. 

Note that the parameter $T_0$ can be absorbed into the H\textsc{i} bias $b_{\rm H\textsc{i}}$ \citep{power_spectrum_equation}. Hence, in real observations, we can actually constrain the parameter product $\Omega_{\rm H\textsc{i}}b_{\rm H\textsc{i}}b_{g}r_{{\rm H\textsc{i}},g}$ using the cross power spectrum $P_{\times}(k)$, if the matter power spectrum $P_{\rm m}(k)$ is known. Furthermore, the constraint on $\Omega_{\rm H\textsc{i}}b_{\rm H\textsc{i}}r_{{\rm H\textsc{i}},g}$ can be achieved, if $b_g$ can be properly estimated. Besides, in the ideal case,  if the auto-correlation of H\textsc{i} intensity mapping can be detected at the same time, the correlation coefficient $r_{{\rm H\textsc{i}},g}$ can be constrained by $P_{\times}/(P_{g}P_{T})$. Unfortunately, due to inevitable foreground residual, experiments which aim at the auto-correlation of H\textsc{i} intensity mapping have not achieved any convincing result so far. Here we assume the H\textsc{i} auto-correlation is unreachable, so that the cosmological parameter we can constrain by cross-correlation power spectrum is basically limited to $\Omega_{\rm H\textsc{i}}b_{\rm H\textsc{i}}b_{g}r_{{\rm H\textsc{i}},g}$ and $\Omega_{\rm H\textsc{i}}b_{\rm H\textsc{i}}r_{{\rm H\textsc{i}},g}$ if $b_g$ can be derived in a galaxy survey.

Since the matter power spectrum could be accurately calculated by a cosmological model, e.g. $\rm \Lambda CDM$, and assuming its uncertainty is small enough that it can be neglected compared to the H\textsc{i} parameters, the theoretical matter power spectrum could be a proper choice for $P_{\rm m}(k)$. We use CAMB \citep{CAMB} to generate a non-linear matter power spectrum $P_{\rm m}(k)$ at $z=0.5$. Here the values of cosmological parameters in CAMB are set to be the same as those in the MultiDark simulation \citep{MultiDark_simulations}, and we can adopt the values obtained from real cosmological observations in the real H\textsc{i} intensity mapping surveys.

In addition, $b_{g}$ can be estimated from galaxy auto power spectrum as indicated in Equation~(\ref{eq:Pg}), and the galaxy auto power spectrum derived from the mock data of CSST galaxy survey is displayed in Figure~\ref{fig:galaxy_power_spectrum}. With the help of the theoretical matter power spectrum, the corresponding $b_{g}$ can be estimated as a function of wavenumber $k$ as shown in Figure~\ref{fig:bg}.
As can be seen, the value of $b_{g}$ presents an increasing trend as the scale gets smaller. This is reasonable since the baryonic matter has more complicated physical mechanisms at smaller scales, like outflow feedback of galaxy and star formation process, which can significantly affect the density fluctuations.
On the other hand, at linear scales, $b_{g}$ is shown to be a constant. We set the scale range to be $k<0.3\, h\mathrm{Mpc^{-1}}$ \citep{MeerKAT&WiggleZ,DFR}, and calculate the average value of $b_{g}$ from the power spectrum. The average value and error of $b_{g}$ are shown in Table~\ref{cosmological parameters}.

In Figure~\ref{fig:Obbr_Obr}, we show the constraint results of $\Omega_{\rm H\textsc{i}}b_{\rm H\textsc{i}}b_{g}r_{{\rm H\textsc{i}},g}$ and $\Omega_{\rm H\textsc{i}}b_{\rm H\textsc{i}}r_{{\rm H\textsc{i}},g}$ as a function of wavenumber $k$. Similar to $b_g$, we derive the average values of these two parameter products in the linear scales with $k<0.3\, h^{-1}\rm{Mpc}$, and the results are listed in Table~\ref{cosmological parameters}. We find that the errors are $\sim1\%$ of the average values, which mean the precision of our future MeerKAT-CSST survey can be one order of magnitude more accurate than the present experimental result \citep{GBT&Deep2,GBT&WiggleZ,GBT&eBOSS}. Furthermore, if assuming $r_{{\rm H\textsc{i}},g}$ is close to 1 at linear scales, which is supported by the simulation results \citep[e.g. see][and our simulation gives $r_{{\rm H\textsc{i}},g}\simeq0.94$ at $k<0.3\, {\rm Mpc^{-1}}h$]{DFR}, the constraint on $\Omega_{\rm H\textsc{i}}b_{\rm H\textsc{i}}$ can be accurately obtained. These indicate that the cross-correlation method is powerful in LIM surveys for cosmic neutral hydrogen, galaxy evolution and cosmological studies.



\section{Summary and Discussion}\label{conclusion and discussion}
H\textsc{i} intensity mapping is a promising technique of cosmological detection. Although due to strong foreground contamination, H\textsc{i} intensity mapping auto-correlation detection is still facing great challenges at the present stage, H\textsc{i}-galaxy cross-correlation could be easier to achieve by extracting H\textsc{i} and cosmological information with the help of galaxy surveys. 
In this work, we have investigated the cross-correlation of MeerKAT H\textsc{i} intensity mapping and CSST spectroscopic galaxy survey at $z\simeq0.5$.

We first generate mock H\textsc{i} intensity maps of MeerKAT and galaxy catalog of CSST spectroscopic galaxy survey using SMDPL $N$-body simulations at $z=0.5$. The system noise and foreground emission are also taken into account when making the sky map. The voxel of simulation box is divided according to the the frequency resolution of $L$-band and beam size of MeerKAT single-dish mode. We constructed the H\textsc{i} model to transform the dark matter distribution in the simulation snapshot into H\textsc{i} distribution. Then we calculated the H\textsc{i} brightness temperature of each voxel to get H\textsc{i} intensity maps. 
The galaxy survey catalog is generated based on the NFW profile and HOD model, and the galaxy luminosity is derived by galaxy luminosity-halo mass relation. Then we filter the galaxies with the magnitude limit of CSST spectroscopic survey to produce the mock galaxy data.

We apply the PCA/SVD algorithm to remove the foregrounds in MeerKAT H\textsc{i} intensity mapping, and cross-correlate the residual intensity map with the corresponding CSST galaxy map. The signal compensation is employed to solve the signal loss caused by the foreground removal process, and we construct the transfer function to compensate the cross power spectrum. After compensation, we derive the H\textsc{i}-galaxy cross power spectrum for constraining cosmological parameters.

We constrain the parameter products $\Omega_{\rm H\textsc{i}}b_{\rm H\textsc{i}}b_{g}r_{{\rm H\textsc{i}},g}$ and $\Omega_{\rm H\textsc{i}}b_{\rm H\textsc{i}}r_{{\rm H\textsc{i}},g}$ using the cross power spectrum, and find that the constraint accuracy can achieve $\sim 1\%$, which is one order of magnitude higher than the current result.
Besides, note that our simulation is in a $\sim300$ $\mathrm{deg}^2$ survey area, and in the future, the MeerKAT-CSST detection of H\textsc{i}-galaxy correlation can be performed on a much larger survey area of $5000\, \mathrm{deg}^2$. Then the constraint accuracy can be reduced to $<0.3\%$ level. This indicates that the cross-correlation of MeerKAT H\textsc{i} intensity mapping and CSST galaxy survey is powerful of exploring the property of cosmic neutral hydrogen and the evolution of galaxies and our Universe.

We also note that some assumptions in the current work still may be too simple which can affect the prediction of the results. For example, the frequency dependency of the beam size may contaminate the data, which will be hard to be removed by the PCA/SVD foreground removal process. Moreover, the non-flat sky effect also should be seriously included, especially for the 5000 deg$^2$ MeerKAT-CSST joint analysis in the future. Other issues, such as the simple HOD model and systematics used in the work, probably need to be considered more carefully with more powerful simulations and precise H\textsc{i} model in future work.

\begin{acknowledgements}
Y.J. and Y.G. acknowledge the support of 2020SKA0110402, MOST-2018YFE0120800, National Key R\&D Program of China No.2022YFF0503404, and the National Natural Science Foundation of China (NSFC, Grant Nos. 11822305, 11773031 and 11633004). 
X.L.C. acknowledges support of the National Natural Sciences Foundation of China (NSFC, Grant Nos. 11473044, 11973047), and the Chinese Academy of Science grants QYZDJ-SSW-SLH017, XDB 23040100, XDA15020200. 
Y.Z.M. is supported by the National Research Foundation of South Africa under grant No.150580, No. 120385 and No. 120378, NITheCS program "New Insights into Astrophysics and Cosmology with Theoretical Models confronting Observational Data".
This work is also supported by the science research grants from the China Manned Space Project with NO.CMS-CSST-2021-B01 and CMS-CSST-2021-A01.
\end{acknowledgements}

\bibliographystyle{raa}
\bibliography{bib}

\end{document}